\documentclass[fleqn,usenatbib]{mnras}

\usepackage{newtxtext,newtxmath}

\usepackage[T1]{fontenc}

\DeclareRobustCommand{\VAN}[3]{#2}
\let\VANthebibliography\thebibliography
\def\thebibliography{\DeclareRobustCommand{\VAN}[3]{##3}\VANthebibliography}

\usepackage{graphicx}	
\usepackage{amsmath}	
\usepackage{lineno}

\usepackage[utf8]{inputenc}
\usepackage{longtable}
\usepackage{graphicx}

\usepackage{mathtools}
\usepackage{paralist}
\usepackage{float}

\usepackage{rotating,booktabs}

\usepackage{subcaption}

 \usepackage{euscript}
 
 \usepackage{xcolor} 
 
 \usepackage{soul}

\title[Elongated Asteroids with 3D Mass Distribution]{A Particle-Linkage Model for Elongated Asteroids with Three-Dimensional Mass Distribution}

\author[Santos, L. B. T. et al.]{
L. B. T. Santos$^{1}$\thanks{E-mail: leonardobarbosat@hotmail.com},
L. O. Marchi$^{1}$,
S. Aljbaae$^{1}$,
P. A. Sousa-Silva$^{3}$,
D. M. Sanchez$^{1}$,
\and A. F. B. A. Prado$^{1}$
\\
$^{1}$Division of Space Mechanics and Control. National Institute for Space Research, INPE. S\~ao Paulo, Brazil.\\
$^{2}$S\~ao Paulo State University, UNESP. S\~ao Jo\~ao da Boa Vista, Brazil.
}

\date{Accepted XXX. Received YYY; in original form ZZZ}

\pubyear{2020}

\begin{document}
\label{firstpage}
\pagerange{\pageref{firstpage}--\pageref{lastpage}}
\maketitle

\begin{abstract}
The goal of the present paper is to develop a simplified model to describe the gravitational fields of elongated asteroids. The proposed model consists of representing an elongated asteroid using a triple-particle-linkage system distributed in the three-dimensional space and it is an extension of previous planar models. A non-linear optimisation method is used to determine the parameters of our model, minimising the errors of all the external equilibrium points with respect to the solutions calculated with a more realistic approach, the Mascon model, which are assumed to give the real values of the system.
The model considered in this article is then applied to three real irregular asteroids: 1620 Geographos, 433 Eros, and 243 Ida. The results show that the current triple-particle-linkage three-dimensional model gives better accuracy when compared to the axisymmetric triple-particle-linkage model available in the literature, and provides an advantage in terms of accuracy over the mass point model, while keeping computational time low.
This model is also used to carry out simulations to characterise regions with solutions that remain bounded or that escape from around each asteroid under analysis. We investigated initial inclinations of $0^\circ$ (direct orbits) and $180^\circ$ (retrograde orbits). We considered the gravitational field of the asteroid, the gravitational attraction of the Sun, and the SRP.
Our results are then compared to the results obtained using the Mascon gravitational model, based on the polyhedral shape source. We found good agreement between the two models.

\end{abstract}




\section{Introduction}
\label{introduction}

For the past few decades, space agencies have focused their attention on visiting the small objects that make up the Solar System, sending spacecraft to asteroids and comets and having plans to send more in the future.

Most asteroids and comets have an irregular shape. When planning a space mission to visit these objects, a question is raised: ``how to effectively represent the gravitational field of these small irregular bodies?''\citep{Elipe1, 2003CeMDA..86..131B}. Several mathematical models have been proposed to overcome this difficulty.

Spherical harmonic expansion is often used to model planets, as these celestial bodies have a sphere-like geometrical shape \citep{Elipe1}. However, when the object has an irregular shape, harmonic expansion is no longer convenient and, often, convergence cannot be guaranteed \citep{1999imda.coll..169R, 2018AdSpR..62.3199J, 2017Ap&SS.362..169L}. A model widely used to represent the gravitational field of irregularly shaped objects is the polyhedral model \citep{1994CeMDA..59..253W, 1996CeMDA..65..313W, 2000JGCD...23..466S} which is considered an accurate model and does not present convergence problems \citep{2000JGCD...23..466S}. Due to its high accuracy, the polyhedral model has been widely used in projects of real missions to analyze the dynamics of the spacecraft near asteroids. As a drawback, the polyhedral model generally contains thousands of parameters (vertices and facets) that are necessary to guarantee its high accuracy. Therefore, this model requires a high computational effort and time to perform the simulations. Besides, an investigation regarding the general properties of the dynamics of the spacecraft as a function of the parameters of the model becomes very difficult. It happens because, in the polyhedral model, the parameters used are mixed, producing a mixed influence on the effective gravitational field of irregular bodies.

Simplified models can often be used to approximate the gravitational field of irregular bodies, requiring less computational effort and producing considerable results in a short amount of time. This topic has been addressed in \cite{2015MNRAS.450.3742C} and \cite{2017MNRAS.464.3552A} where the authors applied the Mascon model using the polyhedral data to place the mass particles. As shown by \cite{2017MNRAS.464.3552A}, the maximum difference between the classical polyhedral approach \citep{1994CeMDA..59..253W, 1996CeMDA..65..313W} and the Mascon 8, considering a uniform density, occurs at the location of the equilibrium point represents 0.11 percent of the distance from the centre of the body, which, in our opinion, may be consider satisfactory. That motivated us to use the Mascon approach to validate our model later on in this work.

Besides that, a simplified model can often determine a simple analytical expression for calculating the gravitational field of irregular bodies \citep{2018RAA....18...84Y}. Another advantage of using a simplified model is that we can easily analyse the consequences of a given parameter on the dynamics around asteroids, such as, for example, the stability of the equilibrium points \citep{2015Ap&SS.356...29Z, 2017Ap&SS.362...61B}, the distribution of stable periodic orbits \citep{2017Ap&SS.362..169L}, as well as the permissible hovering regions \cite{2015RAA....15.1571Y, 2016JGCD...39.1223Z}. Moreover, simplified models can be used to assist orbit design \citep{2017Ap&SS.362..229W} and feedback control \citep{2017Ap&SS.362...27Y}. Particular trajectories can be studied under more accurate models when considering the final steps of a real mission, but a general knowledge of the gravity field based on a small number of parameters is very useful in the first stages of mission design.

Thus, several simplified models have been proposed to represent irregular asteroids, such as the ones in \cite{1999imda.coll..169R, 2001CeMDA..81..235R, 2003CeMDA..86..131B, Elipe1}, who investigated the dynamics of a particle under the gravitational field of an asteroid modelled as a straight segment. Other work that also investigated the motion of a particle around small celestial bodies considering a simplified model used a simple planar plate \citep{Blesa}, a homogeneous cube \citep{2011Ap&SS.334..357L}, a triaxial ellipsoid \citep{2006SJADS...5..252G}, a rotating homogeneous cube \citep{2011Ap&SS.333..409L}, a double and finite straight segments \citep{2014Ap&SS.351...87J}, a rotating mass dipole \citep{Chermnykh, qui, Kokoriev}. In 2015, for the first time, \cite{2015Ap&SS.356...29Z} presented a method for adjusting the rotating mass dipole model to a real asteroid. An improvement of this model was carried out considering the oblateness of the primary bodies \citep{2016Ap&SS.361...14Z, 2016Ap&SS.361...15Z, 2017RAA....17....2Z}, and the dipole segment model \citep{2018AJ....155...85Z}. Based on the elongated asteroid binary systems, \cite{2017Ap&SS.362...61B} used the restricted synchronous three-body problem model. These work, among others, showed that it is useful to apply simplified models to identify the main parameters that dominate the motion of a particle around certain asteroid systems.

Inspired by the work of \cite{2015Ap&SS.356...29Z}, \cite{2017Ap&SS.362..169L} proposed the rotating mass tripole model with symmetrical rotation. In the work developed by \cite{2017Ap&SS.362..169L}, it is argued that small convex bodies can be modeled using the mass tripole model. \cite{2017Ap&SS.362..169L} showed that, from five parameters (which are determined with the help of the polyhedral model), it is possible to define the geometric shape and to obtain the physical characteristics of a real asteroid and to obtain its gravitational field. In order to generalize the work of \cite{2017Ap&SS.362..169L}, \cite{2020RMxAA..56..269D} investigate the qualitative dynamics in the vicinity of an asteroid with an convex shape using a rotating tripole model from a semi-analytical study.

The axisymmetric triple-particle-linkage model, proposed by \cite{2017Ap&SS.362..169L}, considers that the points of mass are located in the $xy$ plane of the asteroid (two-dimensional configuration). But we know that asteroids have a spatial mass distribution ($xyz$ axes). Given this, the purpose of this work is to improve the two-dimensional axisymmetric triple-particle-linkage model (2D tripole) maintaining a simplified format, but considering a three-dimensional mathematical modeling, taking another step towards a more realistic scenario. 

We rely on non-linear optimization to find the parameters of our proposed model, as we will see later in this paper. The advantages of this model are described next.

Because the proposed model considers the distribution of mass in three dimensions, this model reproduces the dynamics of the spacecraft more realistically when compared to the two-dimensional axisymmetric rotating mass tripole \citep{2017Ap&SS.362..169L}, and, consequently, the rotating mass dipole model \citep{2015Ap&SS.356...29Z}.

Although this is a simplified model, it is beneficial to carry out a qualitative investigation to analyse the effects of the model's parameters on the orbital dynamics. Qualitative analyses can be used to investigate the effects of some parameters, such as relating the characteristics of the motion around elongated asteroids and the dynamical properties of the body \citep{2015RAA....15.1571Y}, or to analyse the distribution of stable periodic orbits close to the equatorial plane \citep{2017Ap&SS.362..169L}, etc. Furthermore, qualitative analysis has the potential to conduct preliminary real mission design \citep{2018RAA....18...84Y}, leaving detailed studies for a later stage, based on more accurate models that require more computer time, but will study a small number of trajectories.

This paper is organized as follows. Section \ref{EofM} provides the normalized equation of motion of a particle around elongated asteroids that can be modeled as a two/three-dimensional axisymmetric rotating mass tripole. A general equation is provided, from where we can derive the equation of motion for the situations of the planar or spatial mass of the asteroid, in the inertial and in the rotating frame. Subsequently, the methodology adopted to determine the parameters of the simplified model using a nonlinear optimization method is presented in Section \ref{S2}. In Section \ref{S:3}, the proposed methodology and model are applied to three irregular asteroids: 1620 Geographos, 433 Eros and 243 Ida. A comparison of the three-dimensional axisymmetric rotating mass tripole and two-dimensional axisymmetric rotating mass tripole with respect to the Mascon gravity framework using a shaped polyhedral source is performed in Section \ref{Section5}, whose objective is to show the advantage of this new model. In Section \ref{Section6}, we perform numerical simulations where we construct grids of initial conditions, relating semi-major axis and eccentricity, to characterize bounded and unbounded regions around the asteroids 1620 Geographos, 433 Eros and 243 Ida. Finally, Section \ref{conclusion} provides our conclusions.

\section{Equation of Motion}
\label{EofM}

In this section, we present the mathematical model of the restricted four-body problem, which is a basis for the three-dimensional rotating mass tripole model.

\begin{figure}
\centering\includegraphics[width=1\linewidth]{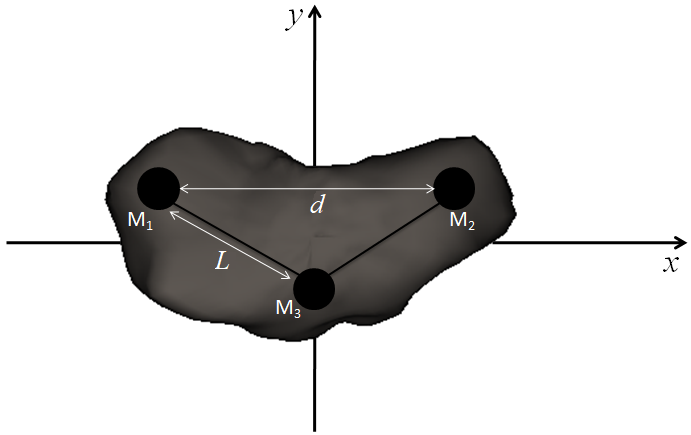}
\caption{The rotating mass tripole in $xy$ plane.}
\label{Fig1}
\end{figure}

Figure \ref{Fig1} shows three points of mass ($M_1$, $M_2$ and $M_3$) arranged inside a body (asteroid or comet) that has an irregular shape. The equations of motion developed in the rotating frame consider the asteroid-spacecraft system, which means that perturbations from other celestial bodies and other external forces are not considered. The rods that connect $M_1$ to $M_3$ and $M_2$ to $M_3$, shown in Fig. \ref{Fig1}, are assumed to have negligible mass and the same length $L$. This length $L$ (in canonical unit) can vary depending on the asteroid (or comet) under study.
The distance between $M_1$ and $M_2$ is denoted by $d$ and it defines one canonical unit (c.u.).

\begin{figure}
\centering\includegraphics[width=1\linewidth]{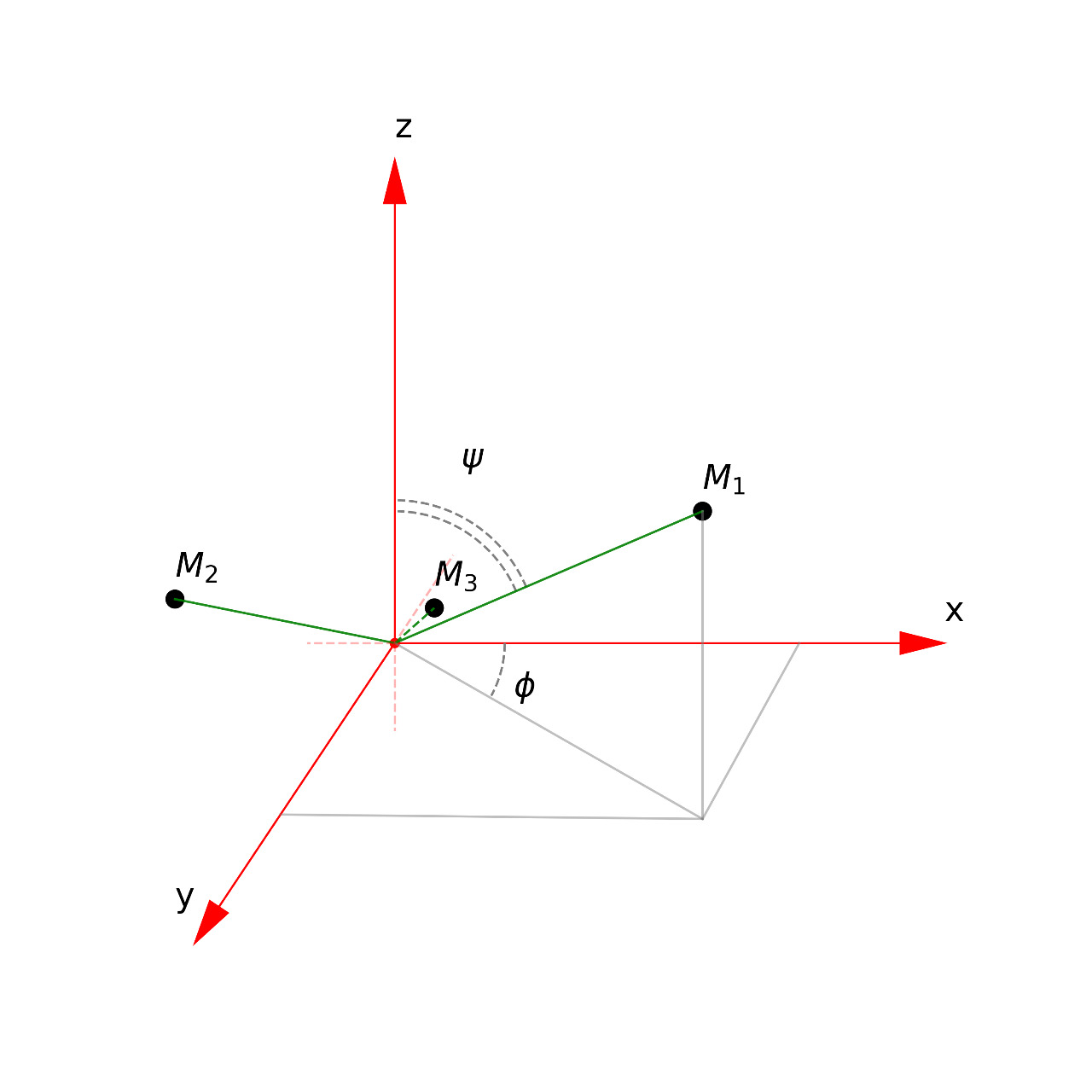}
\caption{Perspective of the positions of bodies $M_1$, $M_2$ and $M_3$ in the three-dimensional space.}
\label{Fig2}
\end{figure}

Note, from Fig. \ref{Fig1}, that it is possible to see the arrangement of the points of mass in the $xy$ plane. Since this study uses a three-dimensional model, we have to analyse the arrangement of these points of mass in space. Figure \ref{Fig2} provides a representative image of the position of $M_1$, $M_2$, and $M_3$ in the $xyz$ space. To determine the locations of the primary bodies, we use two angles, called azimuth ($\Phi$) and elevation angle ($\Psi$). Note that the geometric configuration of the asteroid (or comet) depends on $\Phi$ and $\Psi$. Figure \ref{Fig2} shows the same geometric configuration of the primary bodies. The segment green represents the position of the bodies $M_1$, $M_2$, and $M_3$ with respect to the centre of mass of the asteroid. Note that when $M_1$ and $M_2$ have a positive location on the $y$ axis, $M_3$ has a negative position on the same axis. Likewise, when bodies $M_1$ and $M_2$ have a positive position on the $z$ axis, $M_3$ has a negative position on this same axis, such that the centre of mass of the system remains at the origin of the Cartesian coordinate system (red vectors). From Fig. \ref{Fig2}, we can see that the position of $M_1$ depends on the angles $\Phi$ and $\Psi$. $M_2$ is spatially positioned symmetrically to $M_1$. Although it does not appear in Figure 2, the position of all primary bodies ($M_1$, $M_2$, and $M_3$) depends on the angles $\Phi$ and $\Psi$, as we will see in the equations of motion. We chose not to place all the spatial geometry of the primary bodies with respect to $\Phi$ and $\Psi$ to avoid visual pollution in the figure.

When $\Psi = 90^\circ$ and $\Phi \neq 0^\circ$, we have the particular case of the rotating mass tripole model (two-dimensional). On the other hand, if $\Psi = 90^\circ$ and $\Phi = 0^\circ$, which means that $M_1$, $M_2$, and $M_3$ are aligned on the $x$ axis, there is symmetry with the equatorial plane. We consider that the asteroid's centre of mass is the origin of the reference system ($xyz$).

In this work, we consider that both rods have length $L$, make the same angle $\Phi$, in magnitude, with the $x$-axis and the same angle $\Psi$ with the $z$-axis.

Next, we suppose that a body with negligible mass (spacecraft) is located at P($x$, $y$, $z$). It is convenient to write the equations of motion on a unit scale. For this, let's assume that the distance between $M_1$ and $M_2$ is the unit of length. The time unit is selected such that, the rotational period of the tripole 3D is given by 2\emph{$\uppi$}. It is also assumed that the sum of the masses ($M = m_1+m_2+m_3$) of the primaries is the unit of mass, where $m_1$, $m_2$, and $m_3$ are the masses of $M_1$, $M_2$, and $M_3$, in kilograms, respectively.
The coordinates of the bodies $M_1$, $M_2$ and $M_3$, in canonical units, are, respectively, given

\begin{equation*}
x_1 = -L\cos{\Phi}\sin{\Psi}, 
\end{equation*}
\begin{equation}
\label{eq1}
y_1 = (1-2\mu^*)L\sin{\Phi}, 
\end{equation}
\begin{equation*}
z_1 = L\cos{\Phi}\cos{\Psi},
\end{equation*}

\begin{equation*}
x_2 = L\cos{\Phi}\sin{\Psi}, 
\end{equation*}
\begin{equation}
\label{eq2}
y_2 = (1-2\mu^*)L\sin{\Phi}, 
\end{equation}
\begin{equation*}
z_2 = L\cos{\Phi}\cos{\Psi}, 
\end{equation*}

\begin{equation*}
x_3 = 0, 
\end{equation*}
\begin{equation}
\label{eq3}
y_3 = -2\mu^*L\sin{\Phi}, 
\end{equation}
\begin{equation*}
z_3 = -2\mu^*L\cos{\Phi}\cos{\Psi}/(1-2\mu^*),
\end{equation*}
where $\mu^*$ is the mass ratio defined by
\begin{equation}
\label{eq4}
\mu^* = \frac{m_2}{m_1+m_2+m_3}.
\end{equation}

Note that the distance $d$, in canonical unit, between $M_1$ and $M_2$ depends on $L$, $\Phi$, and $\Psi$, which can be expressed by
\begin{equation}
\label{relation}
x_2 - x_1 = d = 2L\cos{\Phi}\cos{\Psi} = 1.
\end{equation}

In this work, we will use the notation $d$ to represent the distance between $M_1$ and $M_2$ in canonical unit and $d^*$ to represent the distance between $M_1$ and $M_2$ in meters.

Using the canonical units mentioned before, the Lagrangian function of the system is given by:
\begin{multline}
\label{eq5}
\EuScript{L}_{rotating}(x, y, z, \dot{x}, \dot{y}, \dot{z}) = (\dot{x}^2 + \dot{y}^2 + \dot{z}^2)/2+(x\dot{y}-y\dot{x}) + (x^2+y^2)/2  \\
+ k(\mu^*/r_1 + \mu^*/r_2 +(1-2\mu^*)/r_3),
\end{multline}
where
\begin{equation}
\label{eq6}
r_1 = \sqrt{(x-x_1)^2 + (y-y_1)^2 + (z-z_1)^2},
\end{equation}
\begin{equation}
\label{eq7}
r_2 = \sqrt{(x-x_2)^2 + (y-y_2)^2 + (z-z_2)^2},
\end{equation}
\begin{equation}
\label{eq8}
r_3 = \sqrt{(x-x_3)^2 + (y-y_3)^2 + (z-z_3)^2}.
\end{equation}

$k$ is called the force ratio and is given by Eq. \ref{eq9} \citep{2015Ap&SS.356...29Z}
\begin{equation}
\label{eq9}
k = \frac{GM}{\omega^{2}d^{*^3}}.
\end{equation}

Note that $k$ is a dimensionless quantity and depends on the angular velocity vector of the asteroid, denoted by $\omega = [0, 0, \omega]^T$, in radians per second, $M$ is the mass of
the body under study in kilogrammes, $d^*$ is the distance between $M_1$ and $M_2$, in metres and $G$ is the universal gravitation constant, given in the international system of units \citep{2017Ap&SS.362..169L, 2015Ap&SS.356...29Z}.

The Hamiltonian formulation can be obtained from the Lagrangian formulation. The conjugate momenta ($p$) are given by
\begin{equation}
\label{eq10}
p_x = \frac{\partial \EuScript{L}_{rotating}}{\partial \dot x} = \dot x - y,
\end{equation}
\begin{equation}
\label{eq11}
p_y = \frac{\partial \EuScript{L}_{rotating}}{\partial \dot y} = \dot y + x,
\end{equation}
\begin{equation}
\label{eq12}
p_z = \frac{\partial \EuScript{L}_{rotating}}{\partial \dot z} = \dot z.
\end{equation}

Using Legendre transformation ($q, \dot q) \overset{\Lambda}{\rightarrow}(\frac{\partial \EuScript{L}_{rotating}}{\partial \dot q}$) where $q, \dot q$ are coordinates and velocities of the particle, respectively, we get
\begin{equation}
\label{eq13}
\EuScript{H} = \dot x p_x + \dot y p_y + \dot z pz - \EuScript{L}_{rotating}.
\end{equation}

After some simplifications, we obtain the corresponding Hamiltonian
\begin{multline}
\label{eq14}
\EuScript{H}(x, y, z, p_x , p_y , p_z ) = \frac{p_x^2 + p_y^2+ p_z^2}{2} - xp_y + yp_x -\\ k\left(\frac{\mu^*}{r_1}+\frac{\mu^*}{r_2}+\frac{1-2\mu^*}{r_3}\right).
\end{multline}

Note that the Hamiltonian $\EuScript{H}$ is a function of the coordinates and the conjugate momenta $p$ of the spacecraft. Due to the fact that the Hamiltonian is independent of time, it is possible to say that its value is conserved \citep{Benacchio, 1968port.book.....B, Worthington}. $p_x$, $p_y$ and $p_z$ are associated with the spacecraft's conjugate momenta with respect to the $x$-axis, $y$-axis and $z$-axis, respectively.

We define the effective potential by
\begin{equation}
\label{eq15}
\Omega = \frac{x^2 + y^2}{2} + k\left(\frac{\mu^*}{r_1}+\frac{\mu^*}{r_2}+\frac{1-2\mu^*}{r_3}\right).
\end{equation}

From the Hamiltonian function, given by Eq. \ref{eq14}, it is possible to determine the spacecraft's equations of motion in the rotating reference system, given by:
\begin{equation}
\label{eq16}
\ddot x - 2\dot y = \Omega_x,
\end{equation}
\begin{equation}
\label{eq17}
\ddot y + 2\dot x = \Omega_y,
\end{equation}
\begin{equation}
\label{eq18}
\ddot z = \Omega_z.
\end{equation}
where $\Omega_x$ , $\Omega_y$ and $\Omega_z$ is the gradient of the effective potential, as a function of $x$, $y$ and $z$, respectively.

The equations considered in this section, in particular the Hamiltonian function (Eq.(\ref{eq14})) and the equations of motion given by Eqs.(\ref{eq16} - \ref{eq18}), can be used for the two-dimensional (tripole 2D) or three-dimensional (tripole 3D) rotating mass tripole model.

In Section \ref{Section6}, we perform numerical simulations considering the gravitational force of the body in which we want to place a spacecraft to orbit (asteroid or comet), taking into account the solar radiation pressure (SRP) and the gravitational attraction of the Sun. The equations of motion used in the numerical simulation consider an inertial system whose origin of the reference system coincides with the centre of mass of the asteroid. The equations of motion used in the numerical simulation are shown next.

\begin{multline}
\label{eq19}
\ddot x = - \frac{k\mu^*(x' - x_1')}{r'{_1}^3}- \frac{k\mu^*(x' - x_2')}{r'{_2}^3} - \frac{k(1-2\mu^*)(x' - x_3')}{r'{_3}^3} - P_{rad_x}
\\+k_{sun}\frac{M_{sun}}{M_{ast}}\left(\frac{(x_{sun} - x')}{r'^3_{sun}} - \frac{x_{sun}}{r^3_{sun}}\right), 
\end{multline}
\begin{multline}
\label{eq20}
\ddot y = - \frac{k\mu^*(y' - y_1')}{r'{_1}^3}- \frac{k\mu^*(y' - y_2')}{r'{_2}^3} - \frac{k(1-2\mu^*)(y' - y_3')}{r'{_3}^3} - P_{rad_y}
\\+k_{sun}\frac{M_{sun}}{M_{ast}}\left(\frac{(y_{sun} - y')}{r'^3_{sun}} - \frac{y_{sun}}{r^3_{sun}}\right), 
\end{multline}
\begin{multline}
\label{eq21}
\ddot z = - \frac{k\mu^*(z' - z_1')}{r'{_1}^3}- \frac{k\mu^*(z' - z_2')}{r'{_2}^3} - \frac{k(1-2\mu^*)(z' - z_3')}{r'{_3}^3} - P_{rad_z}
\\+k_{sun}\frac{M_{sun}}{M_{ast}}\left(\frac{(z_{sun} - z')}{r'^3_{sun}} - \frac{z_{sun}}{r^3_{sun}}\right),
\end{multline}
where $k_{sun}$ is the force ratio of the Sun, $M_{sun}/M_{ast}$ is the mass of the Sun in canonical unit, $x_{sun}$, $y_{sun}$ and $z_{sun}$ define the position of the Sun on the axis $x$, $y$ and $z$, respectively. In the simulations, we consider that the Sun is in the plane $xy$, consequently, $z_{sun}$ = 0. The positions of the primary bodies $M_1$ , $M_2$ and $M_3$, in the inertial frame, are given by the next equations, respectively.

\begin{equation*}
x'_1 = -L\cos{\Phi}\sin{\Psi}\cos{T} - (1-2\mu^*)L\sin{\Phi}\sin{T}, 
\end{equation*}
\begin{equation}
\label{eq22}
y'_1 = -L\cos{\Phi}\sin{\Psi}\sin{T} + (1-2\mu^*)L\sin{\Phi}\cos{T}, 
\end{equation}
\begin{equation*}
z'_1 = L\cos{\Phi}\cos{\Psi},
\end{equation*}

\begin{equation*}
x'_2 = L\cos{\Phi}\sin{\Psi}\cos{T} - (1-2\mu^*)L\sin{\Phi}\sin{T}, 
\end{equation*}
\begin{equation}
\label{eq23}
y'_2 = L\cos{\Phi}\sin{\Psi}\sin{T} + (1-2\mu^*)L\sin{\Phi}\cos{T}, 
\end{equation}
\begin{equation*}
z'_2 = L\cos{\Phi}\cos{\Psi},
\end{equation*}

\begin{equation*}
x'_3 = 2\mu^*L\sin{\Phi}\sin{T}, 
\end{equation*}
\begin{equation}
\label{eq24}
y'_3 = -2\mu^*L\sin{\Phi}\cos{T}, 
\end{equation}
\begin{equation*}
z'_3 = -2\mu^*L\cos{\Phi}\cos{\Psi}/(1-2\mu^*).
\end{equation*}
where $T$ is the time in canonical unity. The distances from the point mass particle to the bodies $M_1$ , $M_2$ and $M_3$, in the inertial system are, respectively, given by Eqs, \ref{eq25} to \ref{eq27}.
\begin{equation}
\label{eq25}
r_1 = \sqrt{(x'-x'_1)^2 + (y'-y'_1)^2 + (z'-z'_1)^2},
\end{equation}
\begin{equation}
\label{eq26}
r_2 = \sqrt{(x'-x'_2)^2 + (y'-y'_2)^2 + (z'-z'_2)^2},
\end{equation}
\begin{equation}
\label{eq27}
r_3 = \sqrt{(x'-x'_3)^2 + (y'-y'_3)^2 + (z'-z'_3)^2}.
\end{equation}

On the other hand, $r'_{sun}$ and $r_{sun}$ is the Sun-spacecraft distance and Sun-asteroid distance, respectively, given by Eqs. \ref{eq28} and \ref{eq29}.
\begin{equation}
\label{eq28}
r'_{sun} = \sqrt{(x'-x_{sun})^2 + (y'-y_{sun})^2 + (z'-z_{sun})^2},
\end{equation}
\begin{equation}
\label{eq29}
r_{sun} = \sqrt{x_{sun}^2 + y_{sun}^2 + z_{sun}^2},
\end{equation}
where $x_{sun}$, $y_{sun}$ and $z_{sun}$ define the position of the Sun in the inertial system on the $x$, $y$ and $z$ axis, respectively. $P_{rad_x}$, $P_{rad_y}$ and $P_{rad_z}$ represent the components $x$, $y$ and $z$, respectively, of the acceleration due to the SRP \citep{2016AdSpR..57..962P, 2002ApMRv..55B..27M, 2005mcma.book.....B}, given by Eq. \ref{eq30}.

\begin{equation}
\label{eq30}
P_{rad} = C_r\frac{A}{m}P_s\frac{r_0^2}{r'^2_{sun}}\hat{r},
\end{equation}
where $C_r$ is the radiation pressure coefficient. In the simulations we use $C_r$ = 1.5. $m$ is the mass of the spacecraft and $A$ is the area of the straight section of the spacecraft illuminated by the Sun. $P_s$ is the SRP in the vicinity of the Earth-Sun distance and its numerical value is approximately $4.55\times10^{-6}$ N/m$^2$; $r_0$ is the distance between the Sun and the Earth and $\hat{r}$ is the radial unit vector of the Sun with respect to the spacecraft. The value we adopted for the mass of the spacecraft in the simulations was $m$ = 100 kilogrammes (kg) and the area of the solar panel was $A = 1.0$ m$^2$, so $A/m$ = 0.01 m$^2$/kg. Recalling that Equation \ref{eq30} is written in the asteroid-centred inertial frame. 

\section{Determination of the parameters for the symmetrical 2D and 3D triple particle linkage model}
\label{S2}

Two different particle-linkage models were presented in the previous section (2D and 3D tripole). Each particle-linkage model has unknown parameters, and these parameters need to be determined to give a complete set of equations for the model. In this study, considering the simplified model, the total mass and angular velocity of the
asteroids are equivalent to the actual values of those same asteroids. Therefore, the 2D triple-particle linkage model only has four unknown parameters, which are $\Phi$ [rad], $L$ [canonical unit], $k$ [dimensionless], and $\mu^*$ [canonical unit]. The three-dimensional triple-particle linkage model is more complicated. It has five unknown parameters, which are $\Phi$ [rad], $\Psi$ [rad], $L$ [canonical unit], $k$ [dimensionless] and $\mu^*$ [canonical unit].

The unknown parameters are determined using the locations of the external equilibrium points obtained from the Mascon approach, similar to the approach used by \cite{2015Ap&SS.356...29Z, 2016JGCD...39.1223Z, 2018RAA....18...84Y}. The main idea is to find the parameters that generate equilibrium points as close as possible to the ones obtained by using the Mascon model. The locations of the equilibrium points [$\hat{x}_E ,\hat{y}_E,\hat{z}_E$] can be determined as shown by Eq. \ref{eq31}.
This method can be used for asteroids that have external equilibrium points. On the other hand, there are some asteroids do not have external equilibrium points, for example, the asteroid 1998 KY26. For this type of asteroid, the method that minimises the errors of the effective potential or gradient of the effective potential can be used \citep{2018AJ....155...85Z}.

\begin{equation}
\label{eq31}
\Omega_x(\hat{x}_E ,\hat{y}_E,\hat{z}_E) = \Omega_y(\hat{x}_E ,\hat{y}_E,\hat{z}_E) = \Omega_z(\hat{x}_E ,\hat{y}_E,\hat{z}_E) = 0.
\end{equation}

The idea is to find the parameters of the tripole model that minimise the differences of all external equilibrium points between the simplified and the polyhedron model. The method used to determine the parameters for the two/three-dimensional triple-particle-linkage model is described below.
The optimisation variables for the simplified model are $X = [\Phi, \Psi, L, k, \mu^*$].
Before the optimisation, we define the constraints for each variable using the lower limits [$\Phi_{min}, \Psi_{min}, L_{min}, k_{min}, \mu^*_{min}$] and the upper limits [$\Phi_{max}, \Psi_{max}, L_{max}, k_{max}, \mu^*_{max}$] for those parameters. For the nonlinear optimisation problem, the performance index is defined as
\begin{multline}
\label{eq32}
\mathbf{J}(\Phi, \Psi, L, k, \mu^*) = \\\sum_{i = 1}^{n}\sqrt{(\hat{x}_{T_i}d^* - x_{P_i})^2 + (\hat{y}_{T_i}d^* - y_{P_i})^2 + (\hat{z}_{T_i}d^* - z_{P_i})^2},
\end{multline}
where [$\hat{x}_{T_i} ,\hat{y}_{T_i} ,\hat{z}_{T_i}$] are the positions, in canonical units, of the equilibrium points obtained by the simplified models (tripole 2D or 3D). Note that $d$ can be obtained from Eq. \ref{eq9}. On the other hand, [$x_{P_i}, y_{P_i}, z_{P_i}$] represent the locations of the equilibrium points (in metres) obtained by the Mascon model. The index i corresponds to the i-th equilibrium point and, finally, $n$ is the total number of external equilibrium points.
The equality ($c_{eq} = 0$) constraints for tripole 2D and 3D models are shown in Eqs. \ref{eq33} and \ref{eq34}, respectively,
\begin{equation}
\label{eq33}
c_{eq} = \begin{bmatrix}
m_1 + m_2 + m_3 - 1 &\\ 
 L\cos{\Phi} - 1/2& 
\end{bmatrix},
\end{equation}
\begin{equation}
\label{eq34}
c_{eq} = \begin{bmatrix}
m_1 + m_2 + m_3 - 1 &\\ 
 L\cos{\Phi}\sin{\Psi} - 1/2& 
\end{bmatrix}.
\end{equation}

Mathematically, we can write the performance index as a restricted minimisation, as shown in Eq. \ref{eq35}.
\begin{equation}
\label{eq35}
min~\mathbf{J}(\Phi, \Psi, L, k, \mu^*)~such~that~ c_{eq} = 0,
\end{equation}
where $\mathbf{J}(\Phi, \Psi, L, k, \mu^*)$ is a function that returns a scalar value and $c_{eq}$ are functions that return vectors.
Here we used nonlinear optimisation routines developed in Matlab to find the optimal solutions that minimise $\mathbf{J}$. The optimisation problem is defined and solved with a nonlinear programming method (NLP).

\section{Application to realistic elongated asteroids}
\label{S:3}

In this section, we apply the particle-linkage models mentioned in the previous sections to three realistic elongated asteroids. The asteroids are: 1620 Gegraphos, 433 Eros and 243 Ida. The optimisation method is used to determine the parameters of the simplified model, as shown in this section. After that, we demonstrate the improved performance of the symmetric three-dimensional triple-particle-linkage model with respect to the two-dimensional axisymmetric triple-particle-linkage model.

\subsection{Parameters of the sample elongated asteroids}
\label{S:4}

The physical and orbital parameters of the asteroids 1620 Geographos, 433 Eros and 243 Ida were obtained from \cite{2014A&A...568A..43R, 2012P&SS...73...98C} and \cite{2011AJ....141..143B}, respectively, and can be found in Table \ref{table1}. In this study, we also applied the Mascon gravity framework using a shaped polyhedral source to precisely represent the gravitational field of these asteroids \citep{2015MNRAS.450.3742C}. We recommend the readers to review the details of this method in \cite{2013DDA....4420433V, 2017MNRAS.464.3552A}.

\begin{sidewaystable}
\centering

\caption{Physical and Mascon Parameters of 1620 Gegraphos, 433 Eros and 243 Ida.} \label{table1}
\begin{tabular*}{\textwidth}{c @{\extracolsep{\fill}} *{5}{c}}
\hline
\textbf{Asteroid} & \textbf{bulk density} & \textbf{Rotation Period} & \textbf{M}  & \textbf{Verticies $\&$ Facets}\\
   & ($g~cm^{-3}$) & (hours) & (kg)  & \\
   \toprule
\hline
1620 Geographos & 2.15 $\pm$ 0.5 & 5.2233 & 1.65$\times10^{13}$ & 1022 $\&$ 2040\\
433 Eros & 2.67 $\pm$ 0.03 & 5.2702 & 6.69$\times10^{15}$ & 856 $\&$ 1708\\
243 Ida & 2.35 $\pm$ 0.29 & 4.63 & 3.82$\times10^{16}$ & 1022 $\&$ 2040\\
\end{tabular*}

\begin{tabular}{|p{3.5cm}||p{3.5cm}|p{3.5cm}|p{3.5cm}| p{3.5cm}|}
\toprule
\hline
\textbf{Asteroid} & \textbf{$E_1$} (km) & \textbf{$E_2$} (km) & \textbf{$E_3$} (km) & \textbf{$E_4$} (km)\\
\hline
1620 Geographos & [2.6318, 0.1782, 0.0046] & [- 0.0352, 1.9315, 0.0017] & [- 2.6812, 0.2003, -0.0039] & [- 0.0196, -1.9698,
0.0019]\\
433 Eros & [19.1246, -2.5756, 0.1449] & [0.4637,14.7338, - 0.06604] & [-19.6711, - 3.2861, -0.1231]& [-0.4469, -13.991,
-0.0791]\\
243 Ida & [29.7950, -2.5972, 0.6870] & [-0.5688,
25.9245, -0.1113] & [-30.3092, - 1.8763, 0.3958] & [-0.4906, -
25.7145, -0.0993]\\
\hline
\bottomrule
\end{tabular}
\caption{Positions of Equilibrium Points for 1620 Gegraphos, 433 Eros and 243 Ida.}
\label{table2}
\bigskip
\bigskip
\bigskip
\bigskip
\bigskip
\bigskip
\bigskip
\bigskip
\bigskip
\bigskip
\bigskip
\bigskip
\bigskip
\bigskip
\bigskip
\bigskip
\bigskip
\bigskip
\end{sidewaystable}

After adapting the shape of each asteroid in our study according to the procedure presented in \cite{2015MNRAS.452.1316C}, we used the Mascon approach to precisely calculate the positions of the equilibrium points. Table \ref{table2} provides the results.

We recall that we will only consider the external equilibrium points, due to the fact that the internal equilibrium points have no physical meaning.

To perform the simulations, it is necessary to determine the boundary restrictions of each asteroid under study using the particle-linkage model. They are chosen as follows:

\begin{enumerate}
\item 3D triple-particle-linkage model:

For each asteroid, [$\Phi_{min}$, $\Phi_{max}$] are set to [-0.5, 0.5] rad, [$\Psi_{min}$, $\Psi_{max}$] are set to [1.39626, 1.91986], [$L_{min}$, $L_{max}$] are set to [0, 2], [$k_{min}$, $k_{max}$] are set to [0, 9] and [$\mu^*_{min}$, $\mu^*_{max}$] are set to [0.001, 0.999]

\item 2D triple-particle-linkage model:

For each asteroid, [$\Phi_{min}$, $\Phi_{max}$] are set to [-0.5, 0.5] rad, [$L_{min}$, $L_{max}$] are set to [0, 2], [$k_{min}$, $k_{max}$] are set to [0, 9] and [$\mu^*_{min}$, $\mu^*_{max}$] are set to [0.001, 0.999]
\end{enumerate}

The boundary constraints of the geometric parameters $\Phi$, $\Psi$, $L$, are determined based on the actual shapes of the asteroids. Due to the geometric shape of the body and the information we obtain from the equilibrium points from the high fidelity model, we observed that the equilibrium points $E_1$ and $E_3$ do not have high values on the $y$ axis, making us to consider the azimuth angle varying from -$30^\circ$ to + $30^\circ$. Observing the values of the equilibrium points on the $z$ axis, it is possible to notice that these asteroids do not have a very high mass distribution on the $z$ axis, therefore, we consider the elevation angle varying between -$20^\circ$ and + $20^\circ$. Finally, knowing that, by the theoretical definition of our study, the distance between $M_1$ and $M_2$ is one canonical unit, we concluded that the value of $L$ is not much above the numerical value.

The boundary constraints of the physical parameters, $k$ and $\mu^*$, were determined as follows. Through the articles in the literature that use the particle-linkage model \citep{2017Ap&SS.362..169L, 2018RAA....18...84Y, 2015Ap&SS.356...29Z}, we observed that the value of $k$ found in asteroids varies between 0.4 and 9, so we consider an initial guess between 0 and 9. On the other hand, the contour restrictions of the masses were defined based on their theoretical intervals.

After the optimisation has been carried out, if the value of one of the parameters is close to the contour limit, we modify its range and carry out the optimisation again. This procedure is carried out until the values of the optimised parameters are located between the specified limits. The initial guesses for the 2D and 3D model, were chosen as follows:

\begin{enumerate}
\item 3D triple-particle-linkage model:

For 1620 Geographos is [$\Phi$, $\Psi$, $L$, $k$ , $\mu^*$] = [0.36, 1.5, 1, 0.2, 0.24], 433 Eros is [$\Phi$, $\Psi$, $L$, $k$ , $\mu^*$] = [0.3, 1.5, 0.5, 0.2, 0.28] and for 243 Ida is [$\Phi$, $\Psi$, $L$, $k$ , $\mu^*$] = [0.3, 1.5, 1, 0.4, 0.2].

\item 2D triple-particle-linkage model:

For 1620 Geographos is [$\Phi$, $\Psi$, $L$, $k$ , $\mu^*$] = [0.36, $\pi/2$, 1, 0.2, 0.24], 433 Eros is [$\Phi$, $\Psi$, $L$, $k$ , $\mu^*$] = [0.3, $\pi/2$, 0.5, 0.2, 0.28] and 243 Ida is[$\Phi$, $\Psi$, $L$, $k$ , $\mu^*$] = [0.3, $\pi/2$, 1, 0.4,
0.2].
\end{enumerate}

The optimal values of $\Phi$, $\Psi$, $L$, $k$ and $\mu^*$ parameters obtained using the optimisation method mentioned above are shown in Tables \ref{table4} and \ref{table5}, where we see the performance index $J_1$ and $J_2$. These performance indices are calculated after the optimal parameters $J_0$ are obtained. These performance indexes, $J_1$ and $J_2$, provide the relative errors, maximum and minimum, respectively, for a single equilibrium point and are defined as follows

\begin{multline}
\label{eq36}
\mathbf{J_{max}}(\Phi, \Psi, L, k, \mu^*) = \\ max\left(\frac{\sqrt{(\hat{x}_{T_i}d^* - x_{P_i})^2 + (\hat{y}_{T_i}d^* - y_{P_i})^2 + (\hat{z}_{T_i}d^* - z_{P_i})^2}}{L}\right) \\ \times 100\%,~~~~  i = 1, 2, 3, 4,
\end{multline}

\begin{multline}
\label{eq37}
\mathbf{J_{min}}(\Phi, \Psi, L, k, \mu^*) = \\ min\left(\frac{\sqrt{(\hat{x}_{T_i}d^* - x_{P_i})^2 + (\hat{y}_{T_i}d^* - y_{P_i})^2 + (\hat{z}_{T_i}d^* - z_{P_i})^2}}{L}\right) \\ \times 100\%, ~~~~ i = 1, 2, 3, 4.
\end{multline}

\begin{sidewaystable}
\centering
\bigskip
\bigskip
\bigskip
\bigskip
\bigskip
\bigskip
\bigskip
\bigskip
\bigskip
\bigskip
\bigskip
\bigskip
\bigskip
\bigskip
\bigskip
\bigskip
\bigskip
\bigskip
\caption{Optimisation Results for the tripole 3D model.}
 \label{table4}
\begin{tabular}{|l||c|c|c|c|c|c|c|r|}
\textbf{Asteroid} & $\Phi$ & $\Psi$& $L$ & $k$  & $\mu^*$  & \textbf{J} & \textbf{$J_{max}$} & \textbf{$J_{min}$}\\
 & [deg] & [deg] & [c. u.] & [adim] & [c. u.] & [km] & \% & \%\\
\hline
1620 Geographos & 7.1780 & 89.773 & 0.5039 & 0.2991 &  0.2024 & 0.120& 1.5342 & 0\\
433 Eros & -19.892 & 88.7891 & 0.5318 & 0.5195 &  0.2815 & 1.743 & 3.313 & 0\\
243 Ida & -6.3105 & 87.2621 & 0.5036 & 0.6529 &  0.1636 & 2.005 & 2.975 & 0\\

 \end{tabular}
\bigskip
\centering
\caption{Optimisation Results for the tripole 2D model.} \label{table5}
\begin{tabular}{|l||c|c|c| c|c|c|c|r|}
\textbf{Asteroid} & $\Phi$ & $\Psi$& $L$ & $k$  & $\mu^*$  & \textbf{J} & \textbf{$J$}$_{max}$ & \textbf{$J$}$_{min}$\\
 & [deg] & [deg] & [c. u.] & [adim] & [c. u.] & [km] & \% & \%\\
\hline
1620 Geographos & 7.0884 & 90 & 0.5038 & 0.2843 &  0.1971 & 0.147& 1.348 & 0.373\\
433 Eros & -18.935 & 90 & 0.5286 & 0.4454 &  0.2619 & 2.144 & 3.529 & 1.359\\
243 Ida & -7.3614 & 90 & 0.5041 & 0.7169 &  0.1672 & 5.020 & 7.438 & 3.125\\
\end{tabular}
\end{sidewaystable}

\section{Comparison between the simplified and Mascon model}
\label{Section5}

We have verified that the positions of the equilibrium points in the simplified model compared to the Mascon model are closer, giving the first evidence of the validation of the simplified model. Next, we analyse the classifications of the equilibrium points of the three asteroids studied in this paper and we found that our analysis is consistent with the analysis made using the Mascon model. After that, we compared the relative errors between the potential estimated by the 3D tripole model and the Mascon approach \citep{2015MNRAS.450.3742C}, to justify the simplified model.

\begin{figure}
\centering\includegraphics[width=1\linewidth]{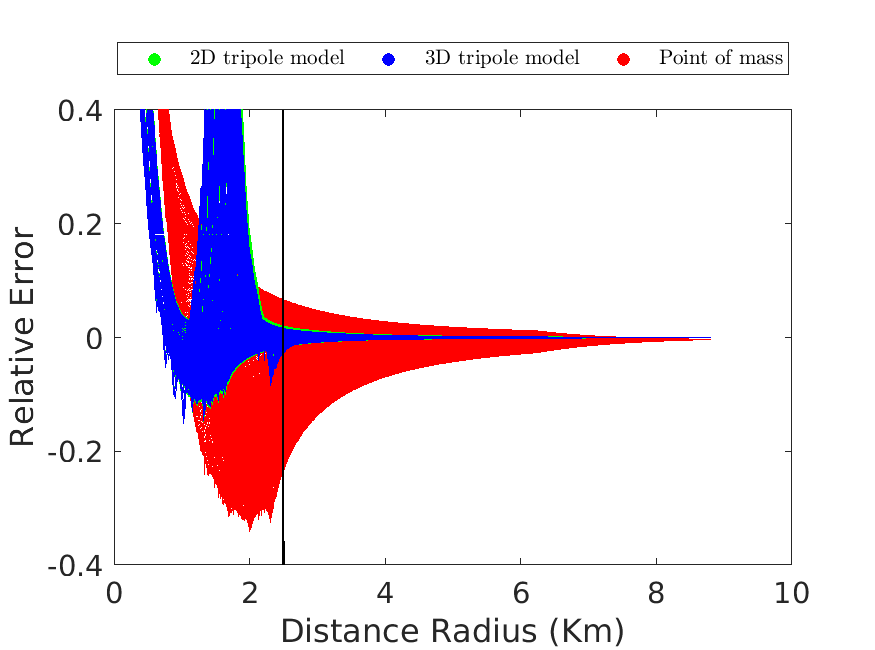}
\caption{Relative error of the gravitational potential for the asteroid 1620 Geographos.}
\label{figure3}
\end{figure}

\begin{figure}
\centering\includegraphics[width=1\linewidth]{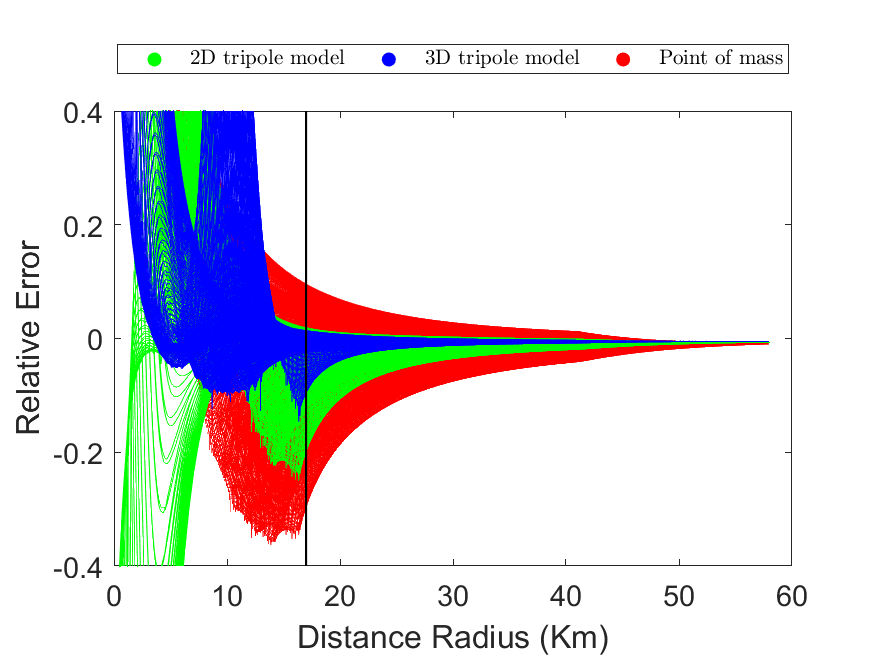}
\caption{Relative error of the gravitational potential for the asteroid 433 Eros.}
\label{figure4}
\end{figure}

\begin{figure}
\centering\includegraphics[width=1\linewidth]{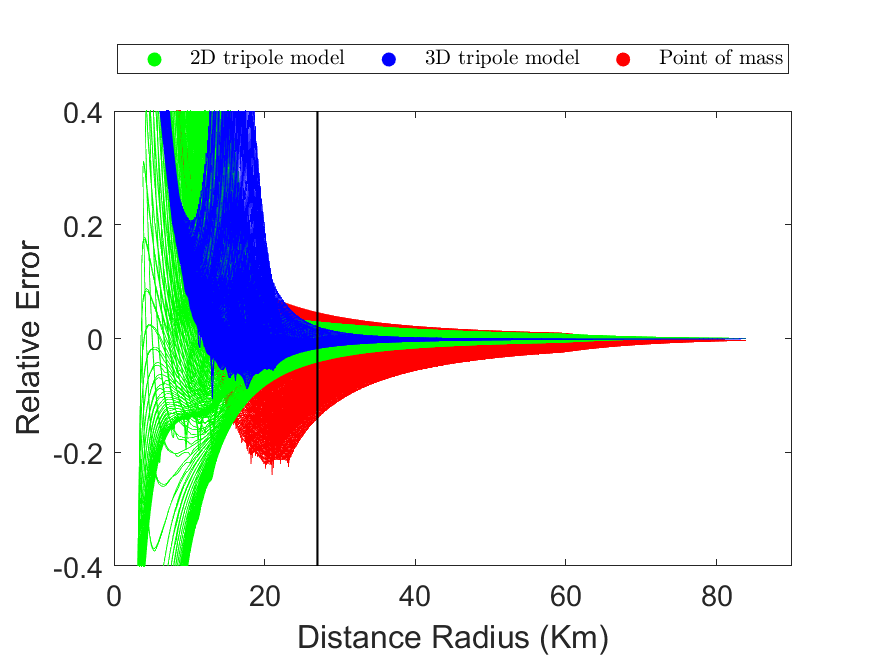}
\caption{Relative error of the gravitational potential for the 243 Ida asteroid.}
\label{figure5}
\end{figure}

Figures \ref{figure3}, \ref{figure4} and \ref{figure5} show the relative errors of the gravitational potentials between the two models (simplified and Mascon model) for the asteroid 1620 Geographos, 433 Eros and 243 Ida, respectively. The black vertical line refers to the surface of the asteroid. The left side of the black vertical line is the relative error of the potential inside the asteroid, and has no meaning for the purpose of this work. On the other hand, outside the asteroid, we observe the relative error of the potential when we consider the asteroid as a mass point (red), 2D tripole model (green) and 3D tripole model (blue).

We see in Fig. \ref{figure3} that, when the spacecraft is close to the asteroid 1620 Geographos (less than 10 km from centre), we cannot consider the asteroid as a point of mass, due to the high relative error between the gravitational potentials. On the other hand, when we model the asteroid as a rotating mass tripole (2D or 3D model), the result shows a good agreement between the models. The 3D tripole model has almost the same accuracy as the 2D tripole model, due to the fact that the mass distribution of this asteroid is predominantly on the $xy$ axis.

Figure \ref{figure4} shows the relative error between the potentials for the asteroid 433 Eros. In contrast to 1620 Geographos, the 3D tripole model is more accurate than the 2D version, when the spacecraft is less than 60 km from the centre of the asteroid.

Finally, Fig. \ref{figure5} shows similar behaviour as in the case of 433 Eros. The 3D tripole model for the asteroid 243 Ida is more accurate than the 2D one.

These results indicate that the 3D tripole model approximates with good accuracy the external potential of some asteroids with elongated bodies. Besides that, it is worth to mention that the tripole model can approximate external potentials for 1030225 points uniformly distributed in the $xy$ plane around an asteroid in about 3 seconds, using an Intel processor of 3.6 GHz. It is about 200 times faster than the Mascon approach shown in \cite{2015MNRAS.450.3742C}, which takes about 10 min. Thus, for a first analysis of a space mission whose objective is to visit an asteroid, it is possible to use the 3D tripole model to obtain preliminary results, since this model is close to the high precision model (Mascon model) and demands well less computational time to perform the simulations.

\section{Numerical investigations}
\label{Section6}

A spacecraft that is close to the surface of an asteroid undergoes complex perturbation, which may lead it to collide with the asteroid or even escape from the domain of its gravitational field. Understanding these perturbations is extremely important to determine regions in space where it is possible to find natural orbits around asteroids. These regions allow a spacecraft to orbit an asteroid for a long time. We will call these regions ``bounded regions''. In contrast, the regions around the asteroids where no natural orbit is possible, will be called ``unbounded regions''.

The present section was developed with the objective of finding possible regions in the vicinity of the asteroids 1620 Geographos, 433 Eros and 243 Ida capable of maintaining a spacecraft for a long time without the need for orbital manoeuvres.

In the simulations performed here, we considered the gravitational field of the asteroid (modelled as a three-dimensional rotating mass tripole) in which it is desired to orbit, the gravitational field of the Sun (modelled as a mass point), and the SRP. The results are shown in the inertial frame with an origin coinciding with the centre of mass of the asteroid under analysis. In the simulations where we take into account the SRP we assume that the area-to-mass ratio is 0.01 $m^2/kg$. The methodology adopted and the results obtained are described next.

\subsection{Numerical simulations methodology} 

To determine the initial conditions of the spacecraft to start the simulations, we initially consider the location of the spacecraft in terms of the Keplerian elements $\mathbf{\kappa} = [a, e, i, \Omega, \omega, f]$. To perform the numerical simulations and build the initial condition grids, we start the orbit with eccentricity 0 and vary it until eccentricity = 0.99, in steps of 0.01. The spacecraft's semi-major axis varies from an initial value $a_i$ to a final value $a_f$, whose values depend on the asteroid that we are analysing. In this study, we investigated direct orbits ($i_0 = 0^\circ$) and retrograde orbits ($i_0 = 180^\circ$). The maximum integration time is 365 days. We use the Runge-Kutta 7/8 method with a time interval, in canonical unit, of 0.01. We consider the initial Keplerian elements as shown in Table \ref{Table6}. The acronym RAAN in the table means Right Ascension of the Ascending Node.

\begin{table}
\centering
\caption{Initial Keplerian elements.}
\label{Table6}
\begin{tabular}{|l| l|}
\hline
Initial semi-major axis [m] & $a_i \in [a_{initial},~ a_{final}]$\\
\hline
Initial eccentricity [adim] & $e_0 \in$ [0, 0.99] \\
\hline
Initial inclination [deg] & $i_0 \in$ [0 and 180] \\
\hline
Initial RAAN [deg] & $\Omega_0 = 0$ \\
\hline
Initial periselene anomaly [deg] & $\omega_0 = 0$ \\
\hline
Initial true anomaly [deg] & $f_0 = 0$ \\
\hline
\end{tabular}
\end{table}

In this initial consideration, we assume that there is only the asteroid that we wish to orbit and the spacecraft. The asteroid, initially, is modelled as a mass point. Based on the initial conditions (the osculating orbit of the spacecraft), we obtain the initial conditions of position ($\mathbf{\rho_0}$ = ($\mathbf{[x'_0, ~y'_0,~z'_0}]^T$) and velocity ($\mathbf{\eta_0}$ = ($\mathbf{[\dot{x}'_0, ~\dot{y}'_0,~\dot{z}'_0}]^T$) of the spacecraft in the Cartesian system, given by $\mathbf{X_0}$ = [$\mathbf{\rho_0},~\mathbf{\eta_0}$], to be used in the equations of motion. $\mathbf{\rho_0}$ and $\mathbf{\eta_0}$ define the state vector of the spacecraft at $t$ = 0 in the restricted problem, where the vector $\mathbf{X_0}$ $\in$ $\mathbb{R}^6$, $\mathbf{\rho_0}$ $\in$ $\mathbb{R}^3$ (initial position) and $\mathbf{\eta_0}$ (initial velocity) $\in$ $\mathbb{R}^3$. The initial conditions obtained, in the Cartesian plane, can be written as $\mathbf{X_0}$ = [$x'_0$ 0, 0, 0, $\pm ~\dot{y}'_0$, 0]$^T$. We assign + $\dot{y}'_0$ when the orbit is direct ($i_0$= 0$^\circ$) and - $\dot{y}'_0$ for retrograde orbits ($i_0$ = 180$^\circ$).

\begin{figure}
\centering\includegraphics[width=1\linewidth]{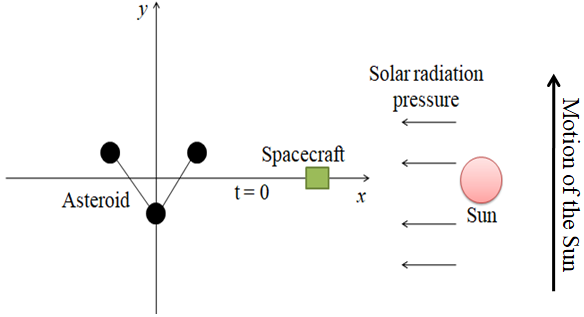}
\caption{Graphic representation of the system studied.}
\label{figure6}
\end{figure}

Once the initial conditions of the spacecraft are found with respect to the asteroid's centre of mass, in the cartesian system, the numerical integration is carried out including the spacecraft, the shape of the asteroid (3D tripole) and the perturbation of the Sun. Regarding the perturbation of the Sun, in some simulations, we consider only the solar gravity perturbation, while in others we consider both, the solar gravity perturbation and the SRP. This is done to represent the situation where the area-to-mass ratio is very small. Initially, ($t$ = 0) the three bodies (Asteroid, spacecraft and Sun) are aligned, as shown in Fig. \ref{figure6} (out of scale). Every simulation is performed in the asteroid-centred inertial frame. To determine the initial conditions of the Sun, we assign the value of the semi-major axis and eccentricity according to references \cite{2014A&A...568A..43R, 2012P&SS...73...98C, 2011AJ....141..143B} for the 1620 Geographos, 433 Eros and 243 Ida, respectively. The other Keplerian elements are considered equal to zero. It means that, initially, the asteroid is located at the Pericentre of its orbit around the Sun. We chose this starting position because the asteroid is in a region close to Earth.
After that, we convert the initial Keplerian elements of the asteroid to Cartesian elements and use them in Equations \ref{eq19}, \ref{eq20} and \ref{eq21}.

During the numerical integrations, we monitor the particles that survived during the integration time, the particles that collided with the asteroid and the particles that were ejected from the system. In the dynamics of the spacecraft, a collision with the asteroid is considered to occur when the spacecraft crosses the disk with a mean radius of the asteroid. We consider an ejection when the spacecraft's distance from the asteroid reaches a value greater than the Hill radius of the asteroid \citep{2000ssd..book.....M}.

The results are presented in graphs that show the lifetime of the spacecraft that survives, for each set of initial conditions, with respect to the semi-major axis and eccentricity of the initial orbit. The regions in which the spacecraft survives for 1 year are called ``bounded regions''. The regions in which the spacecraft survives for a few days are called ``unstable regions''.

\begin{figure}
\begin{subfigure}{0.45\textwidth}
\centering\includegraphics[width=1.0\linewidth, height=5cm]{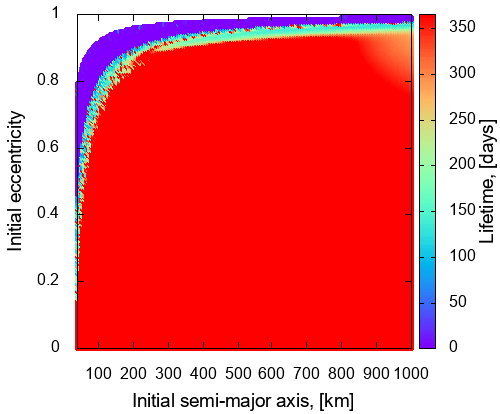} 
\caption{Lifetimes in the regions close to 433 Eros without SRP (direct orbits).}
\label{figure7a}
\end{subfigure}
\begin{subfigure}{0.45\textwidth}
\centering\includegraphics[width=1.0\linewidth, height=5cm]{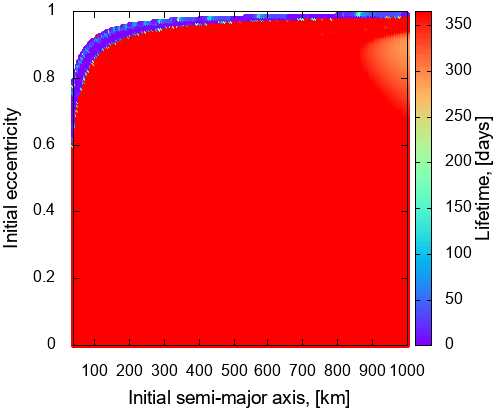}
\caption{Lifetimes in the regions close to 433 Eros  without SRP (retrograde orbits).}
\label{figure7b}
\end{subfigure}

\caption{Plots of $a_0$ versus $e_0$ showing the evolution of the lifetimes in the regions close to 433 Eros asteroid (direct orbits \ref{figure7a} and retrograde orbit \ref{figure7b}) without SRP.}
\label{figure7}
\end{figure}

Figure \ref{figure7a} shows direct orbits and Fig. \ref{figure7b} retrograde orbits taking into account the shape of the asteroid 433 Eros and the solar gravity perturbation. They provide the duration of the orbits as a function of the initial semi-major axis and eccentricity of the particle around 433 Eros. The colour code provides the time the particle remains in orbit for each initial condition. The white area in Fig. \ref{figure7} presents regions where the initial conditions of the spacecraft are inside the asteroid, so it has no physical significance. We investigated the orbital dynamics of a spacecraft with $40 ~\leq ~ a_0 ~ \leq ~ 1000$ km around 433 Eros asteroid. These values of $ a_0 $ were chosen because it is the region where the relative error of the gravitational potential of the rotating triple-particle-linkage three-dimensional model is similar to the potential using the Mascon model, as shown in Fig. \ref{figure4}. Note that the value of $ a_f $ is below the Hill radius ($\approx$ 3270 km) of this asteroid. That allows us to safely neglect the gravitational effects of other celestial bodies. The same reasoning was used to find the range of the semi-major axis of the asteroids studied in this work (1620 Gegraphos and 243 Ida).

It is observed, in Fig. \ref{figure7}, that the solar gravity perturbation has little effect on the dynamics of a spacecraft in the vicinity of asteroid 433 Eros, while the non-spherical gravitational field near the asteroid significantly affects the motion of a particle.

A similar result was found in \cite{Yanshuo}, in which the author showed that solar gravity perturbation is not strong enough to modify families of periodic orbits around the asteroid 433 Eros. In addition, the disturbance of the solar gravity also does not change the stability of the periodic orbits obtained here.

Similar results were found for asteroid 1620 Geographos and 243 Ida. The next results take into account the shape of the asteroid (3D tripole), the solar gravity perturbation and the SRP.

\begin{figure}
\begin{subfigure}{0.45\textwidth}
\centering\includegraphics[width=1.0\linewidth, height=5cm]{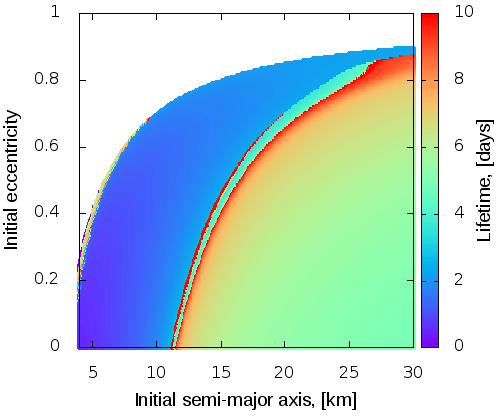} 
\caption{Lifetimes in the regions close to Geographos (direct orbits).}
\label{figure8a}
\end{subfigure}
\begin{subfigure}{0.45\textwidth}
\centering\includegraphics[width=1.0\linewidth, height=5cm]{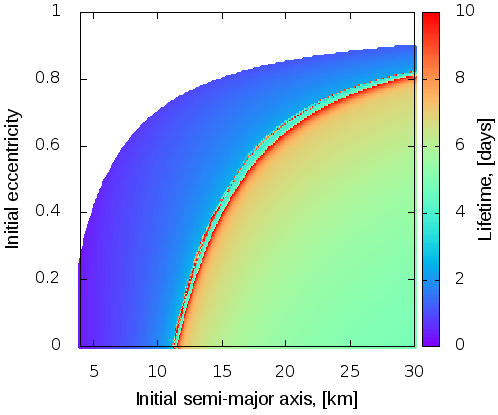}
\caption{Lifetimes in the regions close to Geographos (retrograde orbits).}
\label{figure8b}
\end{subfigure}

\caption{Plots of $a_0$ versus $e_0$ showing the evolution of the lifetimes in the regions close to 1620 Geographos asteroid (direct orbits \ref{figure8a} and retrograde orbits \ref{figure8b}).}
\label{figure8}
\end{figure}

Figure \ref{figure8} provides the initial condition grids in the vicinity of the asteroid 1620 Geographos, where Fig. \ref{figure8a} shows direct orbits and Fig. \ref{figure8b} retrograde orbits. Note that the spacecraft cannot naturally orbit the asteroid 1620 Geographos for a period of 365 days. The maximum duration is around 10 days, but this time is long enough to make observations around the asteroid, so the orbits found here can have practical applications. These results occur due to the small mass of the asteroid 1620 Geographos, which makes the SRP to become significant, making the spacecraft to collide with the asteroid or escape the sphere of its influence.

Note that there is a narrow strip (NS) in Fig. \ref{figure8}, which initially appears at $a$ $\approx$ 11 km and $e$ = 0 and extends to $a$ = 30 km and $e$ $\approx$ 0.9. These regions exist both for direct and retrograde orbits. The maximum survival time of these orbits around the asteroid 1620 Geographos is approximately 10 days. For times longer than this, the spacecraft will escape the system or collide with the asteroid. It is possible to observe that the orbits survive from 4 to 8 days in the regions on the right side of the NS, while the orbits in the region at the left of the NS survive from 0 to 3 days, approximately. The radius of the Pericentre ($r_p = a (1-e))$ coinciding with the position of the narrow strip (NS) are called here $r_{p_{NS}}$.

The regions where $r_p ~ < ~ r_{p_{NS}}$ is close to the asteroid. In this region, the asteroid's gravitational field is strong, making the spacecraft to be attracted by the asteroid or perform a swing-by and quickly escape from the system. On the other hand, in regions where $r_p ~ > ~ r_{p_{NS}}$, the perturbation of the Sun, especially the SRP, dominates the dynamics making the spacecraft to escape from the system in a few days.

\begin{figure}
\begin{subfigure}{0.41\textwidth}
\centering\includegraphics[width=1.0\linewidth, height=6cm]{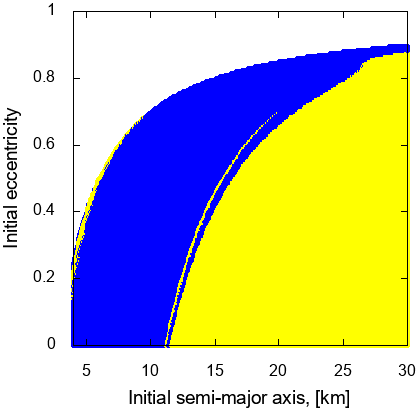} 
\caption{Collision region for direct orbits around Geographos.}
\label{figure9a}
\end{subfigure}
\begin{subfigure}{0.41\textwidth}
\centering\includegraphics[width=1.0\linewidth, height=6cm]{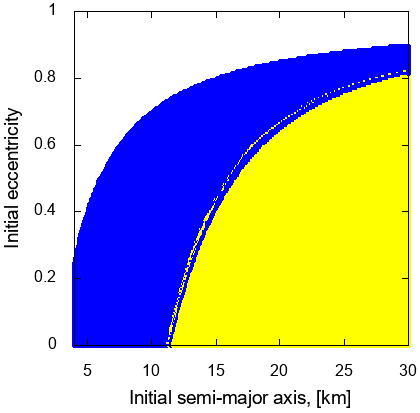}
\caption{Collision region for retrograde orbits around Geographos.}
\label{figure9b}
\end{subfigure}

\caption{Direct orbits \ref{figure9a} and retrograde orbit \ref{figure9b} around Geographos. Collision regions with the asteroid (blue) and regions of ejection from the system (yellow). The white region represents the locations with initial conditions inside the asteroid.}
\label{figure9}
\end{figure}

Figures \ref{figure9a} and \ref{figure9b} provide information about the final destination of the orbits after the integration time shown in Figs. \ref{figure8a} and \ref{figure8b}, respectively. The yellow regions indicate that the spacecraft has escaped the sphere of influence of the asteroid. The blue regions show the initial conditions that lead the spacecraft to collide with the asteroid. Proving our argument, note that, for orbits that start close to the Geographos surface, the spacecraft is captured, ending in collision with the asteroid. Orbits with $r_p ~ > ~ r_{p_{NS}}$ makes the particles to eject from the system. Observe that the boundary between the blue and yellow regions is the narrow strip of 1620 Geographos. It is this boundary that defines whether the orbit will be captured or ejected from the system. Overall, we see that a spacecraft can not orbit the asteroid 1620 Geographos for a long period.

\begin{figure}
\begin{subfigure}{0.45\textwidth}
\centering\includegraphics[width=1.0\linewidth, height=5cm]{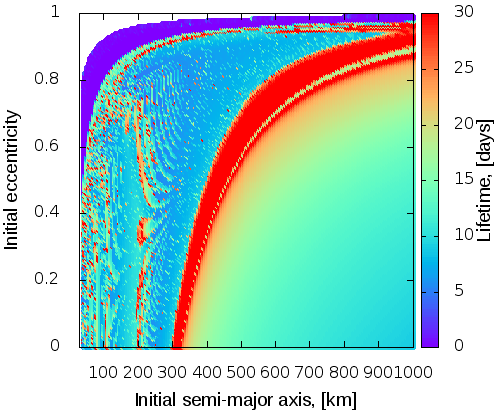} 
\caption{Lifetimes in the regions close to 433 Eros (direct orbits).}
\label{figure10a}
\end{subfigure}
\begin{subfigure}{0.45\textwidth}
\centering\includegraphics[width=1.0\linewidth, height=5cm]{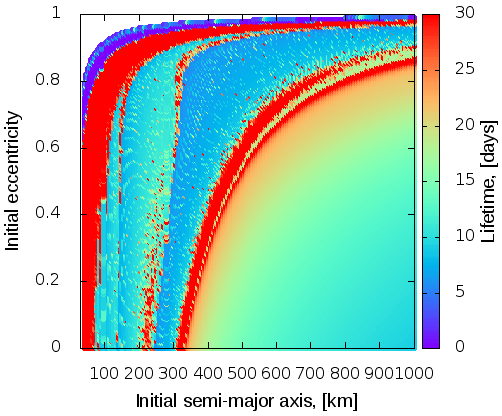}
\caption{Lifetimes in the region closes to 433 Eros (retrograde orbits).}
\label{figure10b}
\end{subfigure}

\caption{Plots of $a_0$ versus $e_0$ showing the evolution of the lifetimes in the regions close to 433 Eros asteroid (direct orbits \ref{figure10a} and retrograde orbits \ref{figure10b}).}
\label{figure10}
\end{figure}

We also make an analysis considering 433 Eros. Some previous studies have analysed the orbital dynamics around this asteroid, neglecting the Sun's disturbances or considering only the solar gravity perturbation, due to the gravitational domain of the asteroid \citep{2000JGCD...23..466S, 2014MNRAS.438.2672C, Yanshuo}. To carry out an investigation complementary to the results available in the literature, we investigated the orbital dynamics of a spacecraft with $40$ km $\leq ~ a_0 ~ \leq ~ 1000$ km, taking into account the effect of the solar gravity and SRP. Our results are presented in Fig. \ref{figure10a} for direct orbits and Fig. \ref{figure10b} for retrograde orbits. We see that there is a narrow strip (starting near $a_0$ $\approx$ 300 km and $e_0 ~ = ~0$) where the spacecraft survives around 30 days. This region corresponds to the $r_{p_{NS}}$ of the 433 Eros.

Note that, for direct orbits, when $r_p$ $<$ $r_{p_{NS}}$, the movement of the particle around the asteroid 433 Eros is unstable due to the shape of 433 Eros and the perturbations from the Sun. Observe that, making a small variation in the initial conditions, it generates completely different trajectories for the particle, reflecting on the lifetime of the orbits and, consequently, in the final destination of the particle. On the other hand, there are regions with solutions that remain bounded close to the asteroid 433 Eros when we consider retrograde orbits (see Fig. \ref{figure10b}).

Space dust most commonly exists around asteroids and planets in direct orbits. On the other hand, retrograde orbits are unlikely to arise naturally. Thus, regions where retrograde orbits survive and direct orbits do not survive are great options for placing a spacecraft, due to the fact that, in these regions, the probability of having space dust is low, decreasing the risk of a spacecraft colliding with some dust particle \citep{2015MNRAS.449.4404A, 2017MNRAS.472.3999A}.

Note that, when we consider the SRP, as presented in this article, the bounded orbits that once existed around 433 Eros cease to exist. This fact shows that the effect of the perturbation of SRP is much larger than the solar gravity perturbation. It happens because the solar gravity perturbation involves direct and indirect terms, that is, the difference between the action of the Sun in the spacecraft and in the asteroid, as shown by Eqs. \ref{eq19} - \ref{eq21}. Those differences are small when the spacecraft gets closer to the asteroid ($\sim$ $10^{-7}$). In contrast, the SRP acts directly on the vehicle, with no indirect term to compensate it, and this makes the effect of SRP to be much larger ($\sim$ $10^{-4}$). There is no doubt that, if we decrease the area mass ratio $A/m$, or if we use the lowest reflectivity coefficient ($C_r$), we can find an increasingly larger regions where the spacecraft remains in orbit around the asteroid for longer times.

\begin{figure}
\begin{subfigure}{0.41\textwidth}
\centering\includegraphics[width=1.0\linewidth, height=6cm]{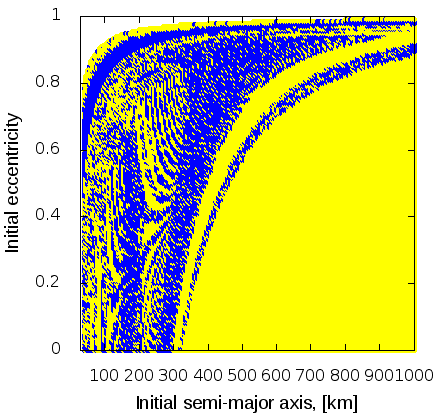} 
\caption{Collision region for direct orbits for 433 Eros.}
\label{figure11a}
\end{subfigure}
\begin{subfigure}{0.41\textwidth}
\centering\includegraphics[width=1.0\linewidth, height=6cm]{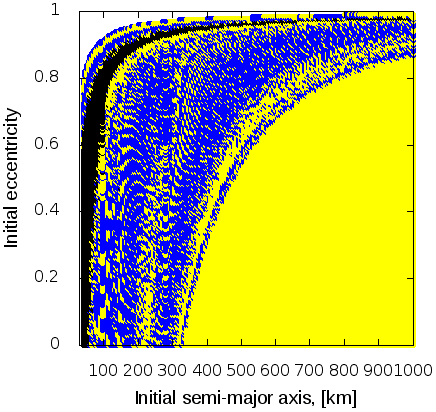}
\caption{Collision region for retrograde orbits for 433 Eros.}
\label{figure11b}
\end{subfigure}

\caption{Direct orbits \ref{figure11a} and retrograde orbits \ref{figure11b} around 433 Eros. Collision region with the asteroid (blue), orbits that survived for 365 days (black) and regions of ejection from the system (yellow). The white region represents the locations with initial conditions inside the asteroid.}
\label{figure11}
\end{figure}

Figures \ref{figure11a} and \ref{figure11b} provide information about the final destination of the orbits after the integration time shown in Figs. \ref{figure10a} and \ref{figure10b}. Note that, for direct orbits, when $r_p~<~r_{p_{NS}}$ km, the spacecraft collides with the asteroid or escapes from the sphere of influence of the asteroid due to the complex dynamics of the movement. For retrograde orbits, we see that there are orbits close to the surface of the asteroid (in black) that survives through numerical integration, that is, that remain bounded.	

Through Figs. \ref{figure10a} and \ref{figure10b}, we see that, when $r_p ~ > ~ r_{p_{NS}}$, for different initial conditions, the spacecraft's lifetime is similar. Note that for small variations of the initial conditions, the spacecraft has approximately the same lifetimes, and probably the same final destination. This statement can be seen in Figs. \ref{figure11a} and \ref{figure11b}. Note that, for $r_p ~ > ~ r_{p_{NS}}$, all orbits are ejected from the system (yellow). This happens because, as we move away from the central body, the disturbing effect of the Sun becomes larger and, consequently, making the spacecraft (or dust particle) to escape the asteroid's sphere of influence.

\begin{figure}
\begin{subfigure}{0.45\textwidth}
\centering\includegraphics[width=1.0\linewidth, height=5cm]{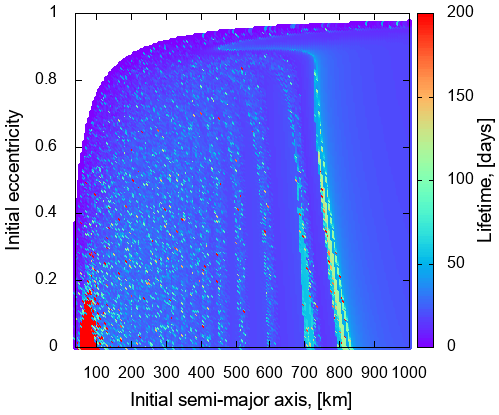} 
\caption{Lifetimes in the regions close to 243 Ida (direct orbits).}
\label{figure12a}
\end{subfigure}
\begin{subfigure}{0.45\textwidth}
\centering\includegraphics[width=1.0\linewidth, height=5cm]{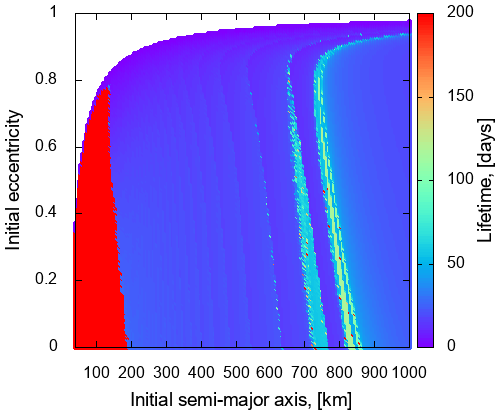}
\caption{Lifetimes in the regions close to 243 Ida (retrograde orbits).}
\label{figure12b}
\end{subfigure}

\caption{Plots of $a_0$ versus $e_0$ showing the evolution of the lifetimes in the regions close to 243 Ida asteroid (direct orbits \ref{figure12a} and retrograde orbits \ref{figure12b}).}
\label{figure12}
\end{figure}

Finally, grids of initial conditions were used to investigate the motion of a spacecraft around 243 Ida, as shown in Figs. \ref{figure12a} and \ref{figure12b}. The orbital elements used to build the grids of initial conditions shown in Figure \ref{figure12} were identical to those used for asteroid 433 Eros. Note that, for direct orbits close to the asteroid ($50$ km $< ~ a_0 ~ <$ 108 km) and low eccentricity, it is possible to observe solutions that remain bounded around asteroid 243 Ida. It was precisely in this region that a moon was observed orbiting the asteroid 243 Ida, called Dactyl, which is a proof that the technique used here works very well.

On the other hand, there is a large region with solutions that remain bounded around 243 Ida when we consider retrograde orbits and $a_0 ~ < ~ 200$ km, as shown in Fig. \ref{figure12b} \citep{2019JSpRo..56.1775S}. Note that this region exist even when the eccentricity is high.

\begin{figure}
\begin{subfigure}{0.4\textwidth}
\centering\includegraphics[width=1.0\linewidth, height=6cm]{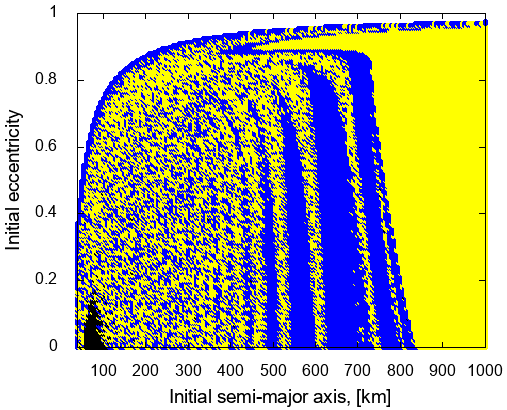} 
\caption{Collision region for direct orbits for 243 Ida.}
\label{figure13a}
\end{subfigure}
\begin{subfigure}{0.4\textwidth}
\centering\includegraphics[width=1.0\linewidth, height=6cm]{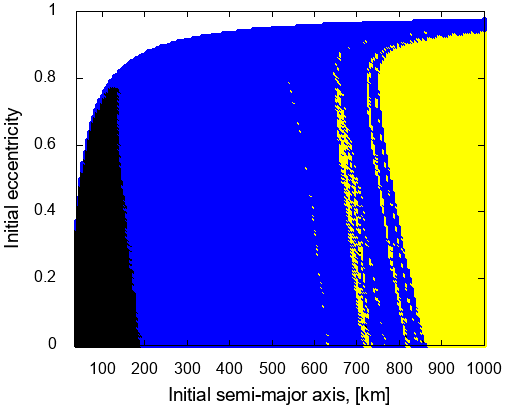}
\caption{Collision region for retrograde orbits for 243 Ida.}
\label{figure13b}
\end{subfigure}

\caption{Direct orbits \ref{figure13a} and retrograde orbit \ref{figure13b} for 243 Ida asteroid. Collision region with the asteroid (blue), orbits that survived for 365 days (black) and regions of ejection from the system (yellow). The white region represents the locations with initial conditions inside the asteroid.}
\label{figure13}
\end{figure}

Figure \ref{figure13} provides information about the final destination of the orbits after the integration time shown in Fig. \ref{figure12}. From the information shown in Fig. \ref{figure13a}, it is possible to conclude that the direct orbits around asteroid 243 Ida are unstable when $a_0~> ~ 120$ km. Note that small variations in the initial conditions make the final destination of the spacecraft to change completely, causing the spacecraft to collide or move away from the asteroid.

For retrograde orbits, note that, when $200$ km  $< ~ a_0 ~< $ 850 km, most of the orbits collide with asteroid 243 Ida. There are few orbits that eject from the system when $a_0 ~ \sim ~ 700$ km. When $a_0 ~ > ~ 850$ km, due to the low gravitational force of 243 Ida, the orbits tend to escape from the sphere of influence of the asteroid.

Similarly to the study made for the asteroid 433 Eros, it is possible to find regions around asteroid 243 Ida in which for direct orbits the regions are unstable, while for retrograde orbits the regions are bounded, those are interesting locations to place a spacecraft, as already explained.

\section{Conclusion}
\label{conclusion}

In this work, we developed a simplified model to represent the gravity field of elongated asteroids (convex or not) with spatial mass distribution, inspired by the existing particle-linkage model. The purpose of this simplified model is to use a three-dimensional axisymmetric triple-particle-linkage model that consists, essentially, of three-point particles distributed in space and two rigid rods with negligible mass. Compared to the mass point model, it was showed that there are ranges of particle-asteroid distances where the accuracy of the proposed model is better, but it still keeps a small processing time compared to high fidelity models. The simplified model was then been applied to the asteroids 1620 Geographos, 433 Eros and 243 Ida. The results indicate that the three-dimensional rotating mass tripole has advantages over the two-dimensional rotating tripole model in terms of accuracy, and get results close to the high fidelity models, using much less computer time.

The results focused on the dynamical study of a spacecraft around the asteroids 1620 Geographos, 433 Eros and 243 Ida, for direct and retrograde orbits, considering an area/mass ratio $A/m ~ = ~ 0.01$ m$^2$/kg. We found solutions that remain bounded in the vicinity of these asteroids. Orbits that remain bounded are important, because they are adequate locations to place a spacecraft. Furthermore, they indicate good places to search for a new member of the asteroid system, as exemplified by the asteroid 243 Ida.

It was observed that no orbit survives for 365 days around the asteroid Geographos when we consider the SRP. On the other hand, we found orbits that remain bounded around the 433 Eros and 243 Ida. Some solutions exist for direct orbits, while other solutions exist for retrograde orbits. Considering the asteroid 433 Eros, the orbits that remain bounded exist only for retrograde orbits. On the other hand, when we investigated the asteroid 243 Ida, we found orbits that remain bounded to both direct and retrograde senses. Determining these regions is important when it comes to astronautical applications, because regions in space where the orbits remain bounded for retrograde orbits and that escape from asteroid for direct orbits are adequate to place a spacecraft, due to the fact that there is a low risk of collisions with dust particles and they provide stability for the orbit, reducing the orbital manoeuvres required to keep the spacecraft orbiting the asteroid. So, simplified models can be used to assist in the pre-designing of real missions.

Finally, It is worth to mention that the small size of our targets allow us to safely neglect the shadowing phenomenon in the SRP. However, Future applications of this model could involve the study of polar orbits considering the effects of the shadowing phenomenon in the SRP.

\section{Acknowledgements}

The authors wish to express their appreciation for the support provided by: grants 140501/2017-7, 150678/2019-3, 422282/2018-9 and 301338/2016-7 from the National Council for Scientific and Technological Development (CNPq); grants 2016/24561-0 and 2016/18418-0, from S\~ao Paulo Research Foundation (FAPESP); grant 88887.374148/2019-00 from the National Council for the Improvement of Higher Education (CAPES) and to the National Institute for Space Research (INPE).

\section*{Data Availability}

The data underlying this article will be shared on reasonable request to the corresponding author



\bibliographystyle{mnras}
\bibliography{example} 





\bsp	
\label{lastpage}
\end{document}